\documentclass[prc,twocolumn,showpacs]{revtex4}

\usepackage{graphicx}

\newcommand{\be}{\begin{equation}}
\newcommand{\ee}{\end{equation}}
\newcommand{\beqn}{\begin{eqnarray}}
\newcommand{\eeqn}{\end{eqnarray}}

\begin{document}

\title{Investigation of the role of shell structure in quasi-fission mass distributions}

\author{D.J. Hinde, R. du Rietz, R.G. Thomas\footnote[1]{
Current address: B.A.R.C.,
Mumbai, India.}, M. Dasgupta, C. Simenel\footnote[2]{
Current address: CEA, Centre de Saclay, IRFU/Service de Physique Nucléaire, F-91191 Gif-sur-Yvette, France}, M.L. Brown, M. Evers, D.H. Luong, L.R. Gasques\footnote[3]{
Current address: Laborat\'{o}rio Pelletron, Instituto de F\'{\i}sica
da Universidade de S\~{a}o Paulo,
05315-970 S\~{a}o Paulo, Brazil.} R. Rafiei\footnote[4]{
Current address: Institute of Materials Engineering, ANSTO, Lucas
Heights, NSW 2234, Australia} and A. Wakhle }

\affiliation{Department of Nuclear Physics, Research School of
Physics and Engineering, The Australian National University,
Canberra,
ACT 0200, Australia\\
}

\date{\today}

\begin{abstract}
Systematic measurements of mass-ratio distributions for fission following collisions of $^{48}$Ti projectiles
with even-even target nuclei from $^{144}$Sm to $^{208}$Pb have been made at sub-barrier energies. They show 
the presence of quasifission, and depend strongly on target nucleus deformation and the fissility of the composite nucleus.
A new framework to analyse systematic mass-ratio measurements allows direct comparison
with the trends expected from shell structure, independent of assumptions or fits. This indicates that quasi-fission mass distributions
show trends consistent with low energy mass-asymmetric fission of the same actinide elements.

\end{abstract}

 \maketitle

The elements found naturally on Earth were formed several billion years ago, many in violent cosmic events such as supernovae. 
The heaviest of these, Thorium (atomic number Z=90) and Uranium (Z=92), have undergone significant decay
since their formation. Their continuing $\alpha$-decay provides much of the energy of volcanos, and the
motion of the continents over the surface of the Earth~\cite{energy}.
From energetics, the ground-states of all atomic nuclei heavier than the Iron/Nickel
elements are in principle unstable to nuclear decay. This can be
understood from the energy change occurring if they were formed by fusion of two lighter nuclei -
the process costs energy, thus the equivalent decay would release energy.
However, hardly any naturally occurring isotopes have measurable decay rates. This is
because the probability of quantum tunneling through
the potential barrier between their compact ground-states and lower energy
configurations is negligible, particularly for decay into two similar mass nuclei, a process called fission. 
However, fission in principle provides a definite limit to the existence of the chemical elements.
For $\beta$-stable nuclei, the liquid drop model
predicts that nuclei with Z$>$115 will have no fission barrier.
However, even heavier elements have been formed in the laboratory~\cite{Ogan118}, with lifetimes $\geq$ms.

This is a result of nuclei with close to integer axis ratios exhibiting ``magic'' numbers
of neutrons and/or protons (closed shells) which have a lower energy than neighboring configurations.
This increases the effective height of the fission barrier, thus providing stability beyond the liquid drop limit.
The shell-stabilization of super-heavy elements was already calculated in 1966~\cite{Sobiczewski66} to occur at neutron number N=184
and Z=114 (more recent calculations~\cite{Bender00} predict stabilization also for Z=120, or even Z=126).
That prediction initiated major experimental efforts to reach this ``holy grail'', to allow testing of nuclear
structure models and relativistic effects in chemistry~\cite{Pyykko77}. Elements up to Z=118 have now been created~\cite{Ogan118},
but forming even heavier isotopes and elements will be more difficult, despite some being predicted to have even longer lifetimes.

The difficulty is that super-heavy elements are formed
in collisions of two massive nuclei, whose large Coulomb repulsion
inhibits fusion.
After contact, the dinuclear system is pictured as diffusing in
shape~\cite{Aritomo09} over the potential energy surface (PES).
The system may reach compact shapes inside the barrier - defined as fusion.
It is generally more likely that following contact it quickly re-separates into fragments
with masses between those of the projectile and target nuclei - a process known as
quasi-fission~\cite{Back85,Toke85} - which inhibits fusion.

The effect of shell structures is central both to the formation of super-heavy elements, as well as to their existence and properties.
Theoretical ideas~\cite{Greiner} and experiments~\cite{Hinde05} suggest that
collisions of doubly closed-shell nuclei give less quasi-fission, thus forming heavy elements with the highest probability.
Indeed, the heaviest elements have been formed~\cite{Ogan118}
in reactions with the doubly closed-shell nucleus $^{48}$Ca.
However, the heaviest practical target nucleus is Cf (Z=98), thus for a $^{48}$Ca projectile, Z is limited to 118.
Shell valleys have also been suggested as a mechanism to form neutron-rich super-heavy isotopes
through inverse quasi-fission~\cite{Zagrebaev} in collisions of two deformed actinide nuclei~\cite{Simenel}.
Shell structures encountered \emph{during} the diffusion are also expected to affect the dynamics~\cite{Aritomo09,ADT1}.
The resulting valleys in the PES are thought to intercept flux~\cite{Zagrebaev} that might otherwise lead to fusion, resulting instead in quasi-fission.
The role of shell structures in fusion dynamics is the most complex and open question, and
is a key element in evaluating prospects~\cite{Zagrebaev} of forming more superheavy isotopes, possibly using neutron-rich radioactive beams.

\begin{figure}[t]
\includegraphics[width=7.5cm]{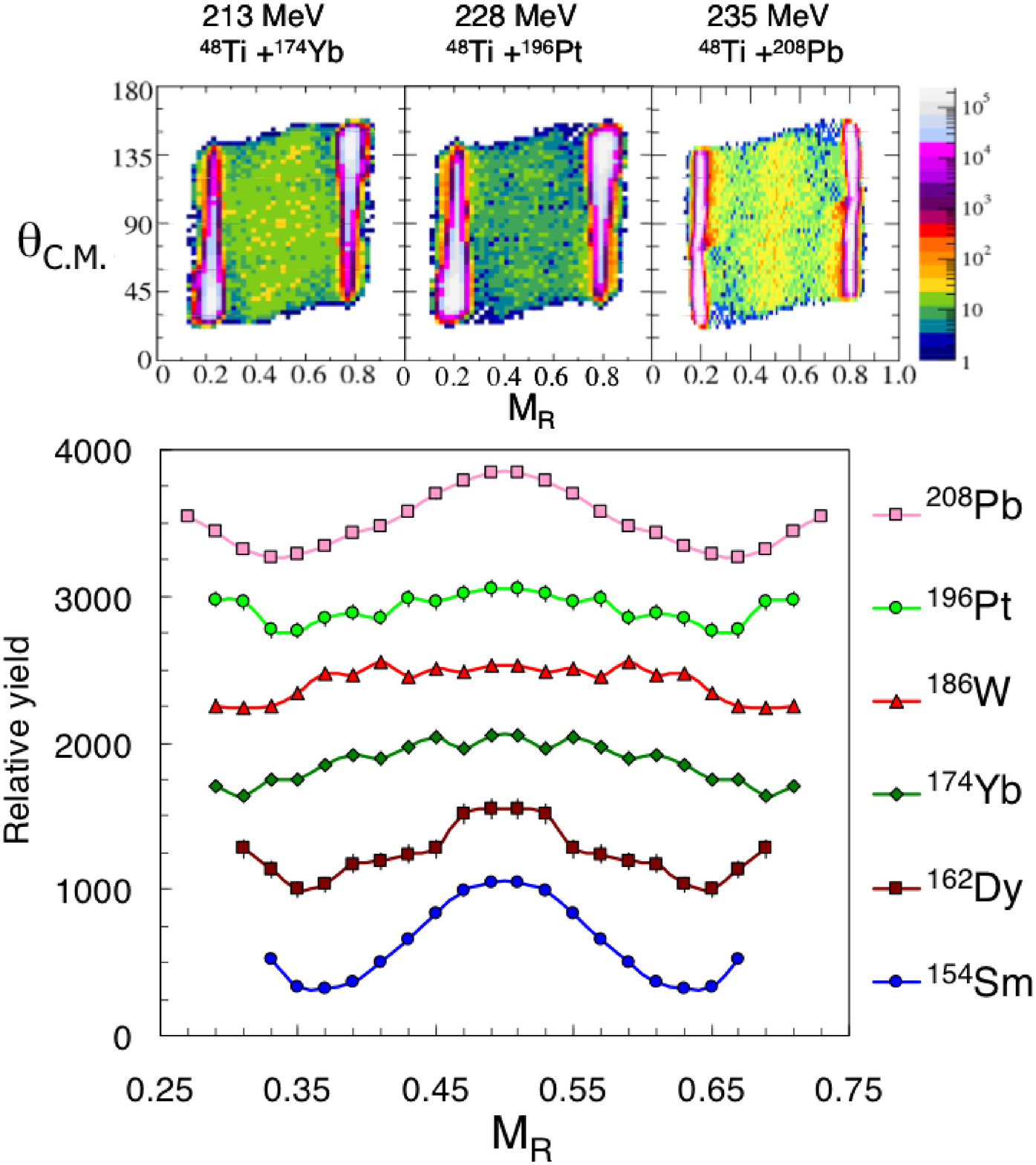}
\caption{\label{Fig2} (Color online) Representative mass-angle distributions (above) and 
projected mass-ratio distributions (below) for the target nuclei indicated. Each projected distribution is normalized to 1050 at the maximum, and
is offset from the lower curve by 500, or 800 for $^{208}$Pb.
}
\end{figure}

Experimental determination of the effects of shell structure, through measuring the characteristics of quasi-fission,
is made challenging by the many factors affecting reaction dynamics.
Even assigning the important neutron and proton numbers that may control quasi-fission mass distributions has up to now involved assumptions.
In this work we propose a new systematic approach. This combines extensive measurements of quasi-fission in reactions forming actinide elements,
with empirical knowledge of the neutron and proton numbers controlling \emph{low energy} fission mass-splits of the same elements.
It makes use of a key feature expected of shell-induced structure in the PES: it will
be present for many fissioning nuclei, at mass-splits where the same fragment is produced.

To investigate experimentally the role of shells in quasi-fission, reactions must be chosen that 
give a substantial quasi-fission yield.
In general, quasi-fission is expected to be more likely the larger the charge product of the
two colliding nuclei, but details of the reaction are also important.
A heavy reaction partner with a large static deformation
results in a range elongations at contact, and thus of entry points into the PES, depending on the orientation. 
If the long axis is aligned with the projectile nucleus, the
dinucleus is elongated at contact, whereas it is more compact if anti-aligned.
The aligned orientation can be selected by choosing a sub-barrier beam energy~\cite{Leigh95,Dasgupta98,Hinde95},
where measurements indeed show the broadest mass distributions, which is strong evidence for quasi-fission
~\cite{Hinde95,Hinde96,Mein97,Hinde02,Thomas08,Rafiei08,Hinde08A}. 
Considering also the expected damping of shell effects with increasing E$_{x}$, 
sub-barrier energies should be optimal for an experimental investigation.

To carry out the measurements, pulsed beams ($\simeq$~1.5~ns FWHM) of $^{48}$Ti in the energy
range 198~--~235~MeV were provided by the ANU 14UD electrostatic
accelerator. Measurements were made at sub-barrier energies, 0.98$\pm$0.01 of the barriers predicted by Swiatecki's model~\cite{Swiatecki05}.
This both increases the probability of quasi-fission and minimizes E$_{x}$,
emphasizing the influence of shell valleys in the PES.
The beams bombarded enriched targets of neutron-rich isotopes of all even-Z targets
from Sm to Pb except Gd; $^{144}$Sm was also included.
The targets were $\sim$50 $\mu$g/cm$^{2}$ in thickness,
evaporated onto $\sim$15~$\mu$g/cm$^{2}$ C backings (facing downstream),
their normal angled at 60$^{\circ}$ to the beam axis. Binary events were
measured in two 28 cm x 36 cm position sensitive multi--wire
proportional counters~\cite{Hinde96}. Located on opposite sides of the beam,
their scattering angle coverage was 5$^{\circ}$~--~80$^{\circ}$ and
50$^{\circ}$~--~125$^{\circ}$, allowing full efficiency detection of all
mass-splits between the projectile and target, for centre-of-mass
angles ($\theta_{c.m.}$) between 40$^{\circ}$ and 140$^{\circ}$.

\begin{figure}[t]
\includegraphics[width=6.5cm]{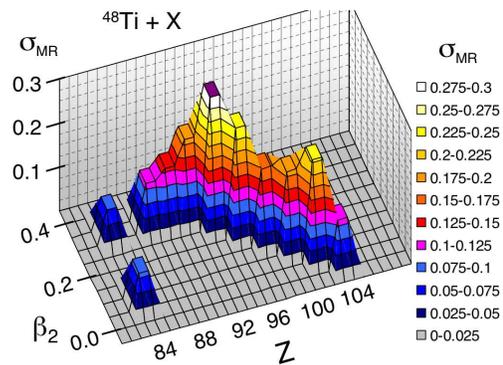}
\caption{\label{Fig3} (Color online)
The width ($\sigma_{MR}$) of the Gaussian best fit to the measured mass-ratio distributions, as a function of
the compound nucleus charge Z and the quadrupole coupling parameter $\beta_{2}$ (see text).
The width for Z=94 ($^{178}$Hf target) is infinite, as the measured distribution
is basically flat. 
}
\end{figure}

\begin{figure*}[t]
\includegraphics[width=17.0cm]{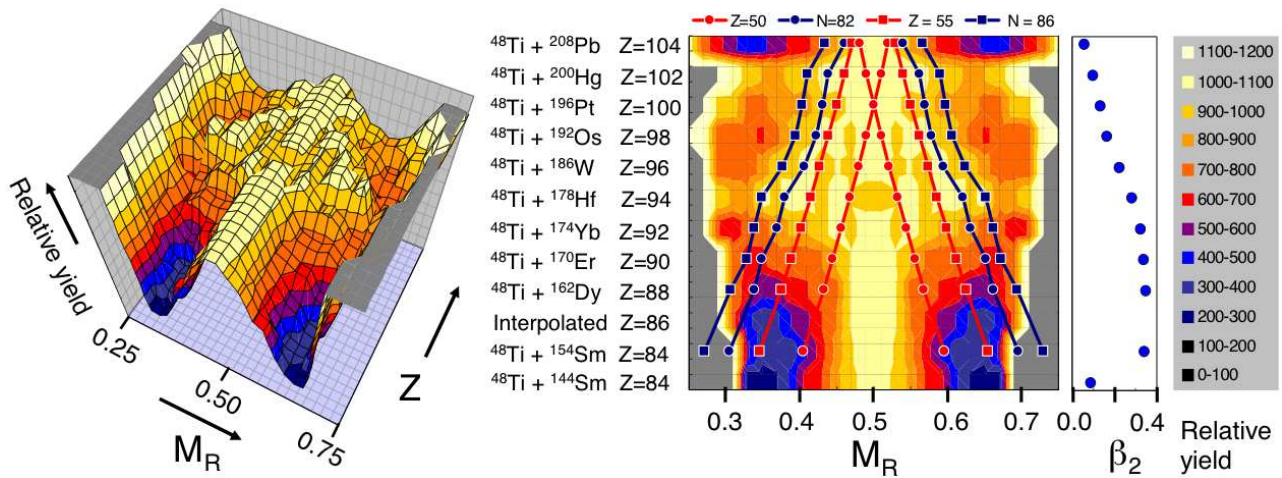}
\caption{\label{Fig4} (Color online)
3-D isometric and 2-D contour representation of the systematics of the \emph{experimental} mass-ratio (M$_{R}$) distributions, as a function of the
atomic number Z of the composite nucleus, for the reactions indicated. The target nucleus $\beta_{2}$ values are plotted
for each reaction. The overlaid joined circles and squares on the 2-D map show the positions of the \emph{empirical}
neutron and proton numbers correlated with the low energy asymmetric fission of the same actinide elements.
}
\end{figure*}

Mass-angle distributions (MAD) were extracted as described in Refs.\cite{Rafiei08,Thomas08,Hinde08A} by
determination of the velocity vector of each coincident particle. The fission mass-ratio M$_{R}$ - the mass 
of one fragment divided by the total mass - was determined event-by-event from the ratio of the
two fragment velocities in the center-of-mass frame~\cite{Hinde96}.
Representative MAD are shown in the upper panels of Fig.\ref{Fig2}. The intense bands at M$_{R}$ $\sim$ 0.2 and 0.8
correspond to elastic and other peripheral collisions, with the fission and quasi-fission
events lying between.

Total mass-ratio distributions were obtained by
projecting the data for 40$^{\circ}$ $<$ $\theta_{c.m.}$ $<$ 140$^{\circ}$.
Six representative distributions of the ten measured are shown in the lower panel.
At M$_{R}$ values more asymmetric than those shown, yields rise rapidly to the elastic peak.
Although some distributions show structure, to present the gross trends
they have been fitted in the range 0.35 $\leq$ M$_{R}$ $\leq$ 0.65 with a Gaussian function.
The resulting $\sigma_{MR}$ values are shown in an isometric histogram in Fig.\ref{Fig3}, as a function of 
both the compound nucleus Z and the absolute value of $\beta_{2}$
coupling the ground-state with the first 2$^{+}$ state of the target nucleus. For $\beta_{2}$ $>$ 0.1, it is reasonable to
associate $\beta_{2}$ with the static deformation. For comparison,
the $\sigma_{MR}$ values for $^{16}$O-induced fusion-fission (where the effect of shells is generally small)
range from $\sim$0.06 to 0.08 for Z=100~\cite{Hinde96}. 

The shape of the wide mass distributions seen for most $^{48}$Ti reactions
must be controlled by two variables: (i) the probability of quasi-fission and (ii) the mass (M$_{R}$) distribution of the quasi-fission.
The narrowest distributions for the $^{48}$Ti reactions (see Fig.3) result from reactions with the
light $^{144,154}$Sm nuclei (forming Z=84), and with the heavy doubly magic $^{208}$Pb nucleus (forming Z=104).
Reactions with Ca projectiles on these targets have previously been associated with small probabilities of quasi-fission~\cite{Itkis07}.
However, quasi-fission is still present for $^{48}$Ti+$^{208}$Pb, as seen in the MAD (Fig.\ref{Fig2}) by the correlation of fission mass with angle~\cite{Hinde08B}.
The $\sigma_{MR}$ values (Fig.\ref{Fig3}) for reactions with the lighter well-deformed ($\beta_{2}$$\sim$0.33) prolate target nuclei 
show a rapid increase in $\sigma_{MR}$ with Z, and thus fissility. However as Z increases beyond 94,
an irregular decrease in $\sigma_{MR}$ occurs, as $\beta_{2}$ reduces.
How these trends correlate with the quasi-fission probability clearly depends on the quasi-fission M$_{R}$ distributions,
which may be affected by shell structure.

To investigate the role of shell structure,
a new representation of systematic experimental M$_{R}$ distributions is introduced, which highlights persistent shell features,
which can be difficult to isolate in limited statistics sub-barrier measurements.
Fig.\ref{Fig4} shows the \emph{experimental} distributions of fission yield (color scale) as a function of 
M$_{R}$ (X-axis) and atomic number Z of the composite system (Y-axis), for all the measurements made in this work. Contour graphs 
are presented both as a 3-D isometric view (left) and as a 2-D map (right).  
As in Fig.\ref{Fig2}, the yield is
normalized to 1050 at the highest point in the distribution, to allow easy visualization of changes of shape with Z.
From $^{154}$Sm to
$^{174}$Yb ($\beta_{2}$ $\sim$0.33) the M$_{R}$ distributions start to show mass-asymmetric shoulders (see also Fig.\ref{Fig2}), 
whose yield increases with system fissility.
These are qualitatively consistent with results for $^{48}$Ca+$^{168}$Er, attributed~\cite{Chizhov03} to quasi-fission. 
The mass distribution is broadest for $^{178}$Hf ($\beta_{2}$ = 0.28) even showing indications of a dip at symmetry.
However, for still heavier targets, having decreasing $\beta_{2}$ but increasing fissility, the shoulders appear to become narrower.
Quasi-fission may be decreasing in probability, and/or shell effects may be modifying the quasi-fission mass distributions. 

\begin{figure}[t]
\includegraphics[width=6.5cm]{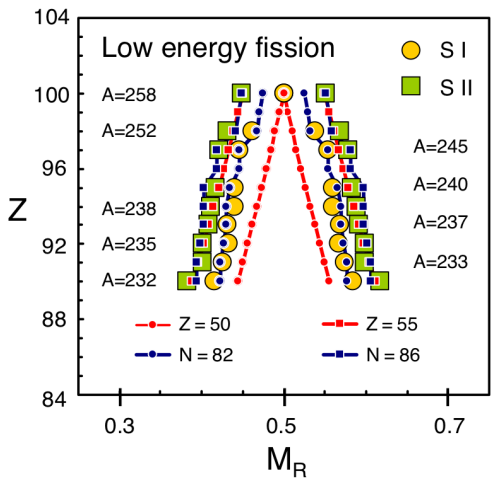}
\caption{\label{Fig1} (Color online) Representation of the systematic trends of
spontaneous and low energy fission mass-ratios for the indicated isotopes of actinide elements, as a function of their atomic number Z.
Large circles and squares represent experimental fission modes, whilst smaller symbols represent trends for
fixed neutron and proton numbers in the fission fragments (see text).
}
\end{figure}

To investigate whether this behavior could be consistent with the effect of shell structures, we treat within the same framework
the systematics for spontaneous and low-energy fission of isotopes of \emph{the same} actinide elements - which exhibit 
shell-driven mass-asymmetric fission, with different mass-split modes~\cite{Pashkevich71,Brosa90}. 
Fig.\ref{Fig1} shows M$_{R}$
as a function of the atomic number Z of the fissioning nucleus.
The large symbols represent the centroids of the empirically determined fission modes for the indicated isotope of each element, taken from
Appendix A of Ref.~\cite{Bockstigel10}, and Ref.~\cite{John71}. 
The ``Standard II'' fission mode (shown by the green squares labeled S~II) generally has the highest yield.
The yellow circles (S I) represent the ``Standard I'' fission mode,
except at Fermium (Z=100), where it represents symmetric fission found for mass number $\geq$ 258, associated with two fragments close to the doubly-magic $^{132}$Sn.
Assuming the N/Z ratios of the fragments are the same as that of the fissioning nucleus (N$^{0}$/Z$^{0}$), for a given particle number Z$^{Shell}$ or N$^{Shell}$ in the fragment, the associated mass ratio M$_{R}^{Shell}$ is given by the ratio Z$^{Shell}$/Z$^{0}$ or N$^{Shell}$/N$^{0}$.
The joined small circles indicate the M$_{R}^{Shell}$ values associated with the spherical closed shells Z=50 and N=82. 
The small squares show the expected trends associated with Z=55 and N=86, the proton and neutron numbers empirically found to be closely associated~\cite{Bockstigel10} with the generally predominant ``Standard II'' mode. 
The systematic behavior of the mass-splits is consistent with the trends expected if fixed proton and/or neutron
numbers in the nascent fragments~\cite{Bockstigel10} are responsible.
It appears that the spherical shells do not play the most significant role in low energy mass asymmetric fission.

To compare with the quasi-fission data, overlaid on the 2-D contour map in Fig.\ref{Fig4} are plotted the M$_{R}^{Shell}$ values for the same shell numbers.
For the measured distributions with Z$\geq$94, it is striking that areas of high yield away from symmetry show a systematic correlation with the
trends of the shell structures. This suggests that
shell structure in the PES does contribute to the observed mass distributions in these quasifission reactions, although not necessarily the spherical shells.
Shell structure in \emph{both} N and Z at the \emph{same} mass-ratio should affect mass distributions most strongly.
In these measurements the empirical Z=55 and N=86 lines are not as close together as in low energy fission, because the isotopes
formed following capture of $^{48}$Ti projectiles are less neutron-rich.
Their occurrence at \emph{different} mass-splits means that we should not necessarily expect to see in these quasi-fission measurements
asymmetric fission peaks identical to those in low energy fission of more neutron-rich isotopes.

This new approach allows investigation and tracking of
neutron and proton shell effects in both low energy fission and quasi-fission.
How the effects of shells reinforce each other could be investigated through a systematic study of sub-barrier quasi-fission, 
by varying the composite system neutron number for fixed Z.
If shells in both Z and N occur at the same M$_{R}$, this should result in more distinct peaks in the quasi-fission mass distributions.
Multi-dimensional PES calculations will help to predict optimal reactions for measurements.
Such a study would be a good candidate for an early application of radioactive beams to better understanding superheavy element formation.


The authors acknowledge support from ARC Discovery Grants DP0664077 and DP110102858.


\end{document}